\documentclass[twocolumn,prb,showpacs]{revtex4}
\usepackage{amsmath,amsthm}
\usepackage{epsfig}
\usepackage{graphics}
\usepackage{graphicx}
\usepackage{dcolumn}
\usepackage{bm}

\begin{document}

\title{Correlated metallic state of vanadium dioxide}
\author{M. M. Qazilbash,$^{1,\ast}$ K. S. Burch,$^1$ D. Whisler,$^1$ D. Shrekenhamer,$^1$
B. G. Chae,$^2$ H. T. Kim,$^2$ and D. N. Basov$^1$}
\affiliation{
$^{1}$Physics Department, University of California-San Diego, La Jolla, California 92093, USA \\
$^{2}$IT Convergence and Components Lab, ETRI, Daejeon 305-350,
Korea}

\date{\today}

\begin{abstract}
The metal-insulator transition and unconventional metallic
transport in vanadium dioxide (VO$_2$) are investigated with a
combination of spectroscopic ellipsometry and reflectance
measurements. The data indicates that electronic correlations, not
electron-phonon interactions, govern charge dynamics in the
metallic state of VO$_2$. This study focuses on the frequency and
temperature dependence of the conductivity in the regime of
extremely short mean free path violating the Ioffe-Regel-Mott
limit of metallic transport. The standard quasiparticle picture of
charge conduction is found to be untenable in metallic VO$_2$.

\end{abstract}


\maketitle

\section{Introduction}

Transition metal oxides constitute outstanding systems to explore
strong correlations in solids. A rich variety of phase transitions
and exotic electronic and magnetic ground states in this class of
materials are at the focus of condensed matter
physics.\cite{imada-rmp} One system that has received considerable
attention is vanadium dioxide (VO$_2$) that undergoes a first
order metal-insulator transition (MIT) at $T_c$ $\approx$ 340 K
between a high temperature rutile metallic state and a low
temperature insulating state. The renewed interest in this
particular oxide is, in part, due to unresolved roles of the
structural transformation and of Coulomb interactions in the
MIT.\cite{ allenprl,georges,chae-kim,softxray,okazaki,noh} A
better understanding of the metallic state is required to
elucidate the nature of the MIT.

Yet another remarkable property of VO$_2$ pertains to the
unconventional behavior of the resistivity for
$T>T_c$.\cite{allenprb} The electronic mean free path inferred
from the resistivity data was nearly the same as the lattice
constant, yet resistivity showed linear $T$ dependence without
signs of saturation. In typical metals, the resistivity increases
with increasing temperature but begins to saturate when the mean
free path ($l$) becomes comparable to the lattice constant
($d$).\cite{allencomments} Ioffe and Regel, and subsequently Mott,
have expressed the view that the electron mean free path cannot be
less than the lattice constant, and $l \simeq d$ is commonly
referred to as the Ioffe-Regel-Mott (IRM)
limit.\cite{ioffe,mott,gurvitch} However, in certain exotic
metals, including the metallic phase of VO$_2$ and the high-$T_c$
superconductors, the resistivity continues to increase beyond the
IRM limit.\cite{gunnarsson,hussey} A complete understanding of
this behavior in these so-called ``bad metals"\cite{emery} is
lacking. More importantly, when $l \lesssim d$, the conventional
picture of a quasiparticle with a well-defined momentum undergoing
occasional scattering events breaks down. Such a regime can no
longer be described by the Boltzmann equation, signalling a
breakdown of Fermi liquid theory.\cite{allencomments} Fermi liquid
theory has been remarkably successful in describing the electronic
properties of typical metallic systems. Understanding the
deviations from this theory in exotic metals is a central theme of
modern research.\cite{imada-rmp}

Significant progress has been made recently in the understanding
of resistivity in the regime of short electronic mean free path.
\cite{gunnarsson,hussey,merino} Gunnarsson \emph{et al.}, in
particular, have supplemented early qualitative discussion with
numerical work, including the role of
correlations.\cite{gunnarsson} They devised a new criterion for
resistivity saturation in correlated metals taking advantage of
optical sum rules. The new criterion predicts saturation
resistivities at values exceeding the IRM limit and this is
attributed to the reduction of electronic kinetic energy due to
correlation effects. Thus motivated, we have investigated the
metallic phase of VO$_2$ with infrared/optical spectroscopy over a
wide frequency range and up to high temperatures. Surprisingly,
the optical conductivity shows typical metallic behavior beyond
the IRM limit with the Drude peak centered at zero frequency
persisting at high temperatures. In this work, we discover that
beyond the IRM limit, the linear increase of resistivity of VO$_2$
with temperature is due to the linear increase of scattering rate
of charge carriers. We show that although VO$_2$ violates the IRM
criterion, the data is consistent with a higher value of
saturation resistivity based on Gunnarsson $et. al's.$
model.\cite{gunnarsson} We conclude that electronic correlations,
not electron-phonon interactions, dominate charge dynamics in the
high temperature metallic phase of VO$_2$ and that the correlated
charge carriers are not conventional Landau quasiparticles.

\section{Experiment, Results, and Discussion}

VO$_2$ films of thickness $\approx$ 1000 \AA~were grown on
Al$_2$O$_3$ (sapphire) substrates with the sol-gel method. X-ray
diffraction shows that the films are single phase VO$_2$. The
resistivity decreases by four orders of magnitude upon entering
the metallic state above $T_c$ $\approx$ 340 K. The details of
film growth and characterization are given in Ref.\cite{chae}. The
optical constants were obtained directly from a combination of
ellipsometric measurements in the spectral range 50 meV - 5.5 eV
and near-normal incidence reflectance data in the spectral range 6
meV - 90 meV.\cite{kenny}

\begin{figure}[t]
\epsfig{figure=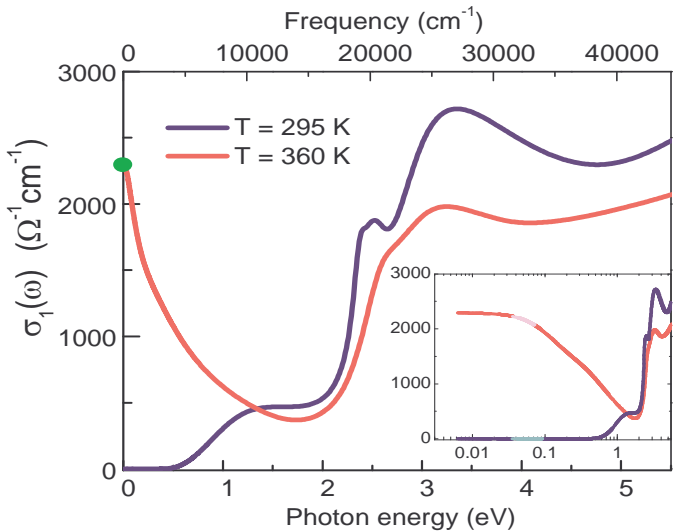,width=90mm,height=70mm}
\caption{(color online) Real
part of the optical conductivity $\sigma_1(\omega)$ of VO$_2$ in
the insulating state ($T$= 295 K) and metallic state ($T$ = 360
K). The solid circle refers to $\sigma_1$($\omega \rightarrow$ 0)
= 2300 $\Omega^{-1}$cm$^{-1}$. Inset: the same data is shown on a
logarithmic scale (in eV). Part of the data with phonons removed
is depicted by lighter shades.\cite{footnote1}} \label{sigma1}
\end{figure}

In Fig.~\ref{sigma1}, we display the real part of the optical
conductivity $\sigma_1(\omega)$ below and above $T_c$. VO$_2$
shows a band-gap of $\approx$ 0.6 eV with a clear threshold in
$\sigma_1(\omega)$ in the insulating state ($T$ = 295 K). The band
gap value is in accord with previous
data,\cite{verleur,noh,okazakioptics} and the presence of a clear
threshold confirms the good quality of sol-gel grown VO$_2$ films.
Three distinct peaks at 1.4 eV, 2.5 eV and 3.4 eV that arise from
inter-band transitions can be discerned in the insulating state.
There is significant rearrangement of the spectrum as VO$_2$
enters the metallic regime ($T$ = 360 K). A rather broad
Drude-like peak emerges at low energies and signifies metallic
behavior. There is a shift of spectral weight from the inter-band
transitions at 2.5 eV and 3.4 eV to the Drude peak. The difference
in the conductivity of the two phases extends beyond the upper
cut-off of our measurements at 5.5 eV. Similar broad energy scales
are involved in the redistribution of spectral weight between the
insulating and metallic states in other correlated electron
materials\cite{imada-rmp} including a close counterpart
V$_2$O$_3$.\cite{rozenberg} Therefore, data in Fig.1 indicates the
presence of electronic correlations in VO$_2$.

A more quantitative approach to assess the importance of
correlations is through their effect on the kinetic energy of the
conduction electrons. If electron-electron interactions are
important, then the measured kinetic energy is expected to be
substantially reduced from the value given by band
theory.\cite{millisreview} This has indeed been observed in both
the hole-doped and electron-doped cuprates.\cite{millisndoped} The
optical conductivity provides us the means to obtain the
expectation value of the kinetic energy of the conduction
electrons in a solid through the partial sum
rule:\cite{gunnarsson}

\begin{equation}
\frac{\omega{_p}{^2}}{8}=\int_{0}^{\omega_c}\sigma_1(\omega)d\omega=-\frac{{\pi}d^2e^2}{6N\Omega\hbar^2}\langle{T_k}\rangle
\label{sumrule}
\end{equation}

Here, $\omega_p$ is the plasma frequency, $d$ is the lattice
constant, $\Omega$ is the volume of a unit cell, $N$ is the number
of unit cells and $\langle{T_K}\rangle$ is the expectation value
of the kinetic-energy operator. Thus the expectation value of the
kinetic energy is simply proportional to the square of the plasma
frequency $\omega_p$. We obtain the plasma frequency by
integrating the area under the intra-band Drude part of the
conductivity up to a cutoff ($\omega_c$) determined by the minimum
in the conductivity at 1.7 eV and find $\omega_p$ = 2.75 eV. The
square of this value is only fifty percent of that obtained from
band theory calculations.\cite{allenprb,kotliar-unpublished} This
result is not particularly sensitive to the choice of the cutoff
and confirms the importance of correlations in the conducting
state of VO$_2$.

\begin{figure}[t]
\epsfig{figure=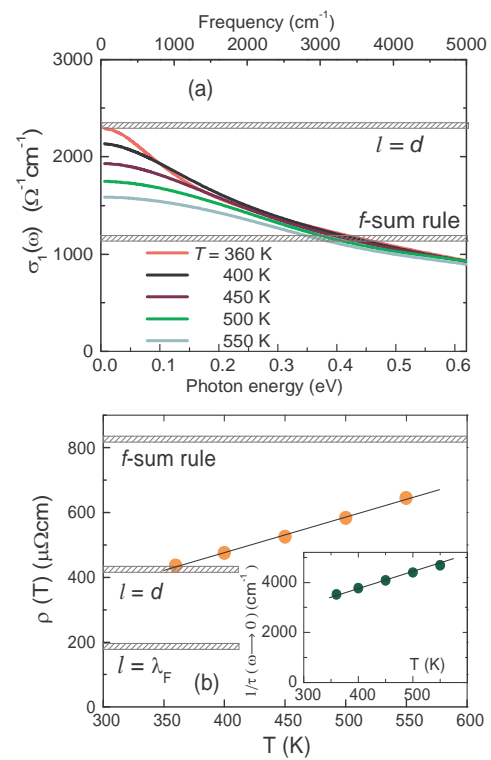,width=65mm,height=100mm}
\caption{(color online) (a)
Frequency dependence of $\sigma_1(\omega)$ of metallic VO$_2$ at
different temperatures. Phonons have been removed in the frequency
range 0.04 - 0.08 eV.\cite{footnote1}(b) Resistivity $\rho(T)$ of
metallic VO$_2$, defined as 1/$\sigma_1$($\omega \rightarrow$ 0,
$T$), is plotted as a function of temperature (circles). Inset:
The scattering rate $1/\tau (\omega \rightarrow 0)$ is plotted as
a function of temperature (circles). The solid lines in panel (b)
and the inset are guides to the eye. The annotated horizontal
lines in panels (a) and (b) depict $dc$ conductivity and
resistivity values at which saturation is expected according to
the criteria mentioned in the text.}\label{sigmaT}
\end{figure}

We investigate further the metallic phase of VO$_2$ at elevated
temperatures. In Fig.~\ref{sigmaT}a, we display the frequency
dependence of $\sigma_1(\omega)$ in the metallic phase of VO$_2$
for successively increasing temperatures. The conductivity
decreases with increasing temperature as coherent spectral weight
is transferred above 0.6 eV. However, the conductivity retains the
Drude-like shape even at the highest temperature attained. In
Fig.\ref{sigmaT}b we plot the ``optical" resistivity
$\rho(T)=1/\sigma_1(\omega \rightarrow 0, T)$. The resistivity
continues to increase almost linearly with temperature up to the
upper limit of our measurements at  $T$ = 550 K without any sign
of saturation. This is consistent with lack of saturation seen in
$dc$ resistivity measurements.\cite{allenprb}

\begin{figure*}[t]
\epsfig{figure=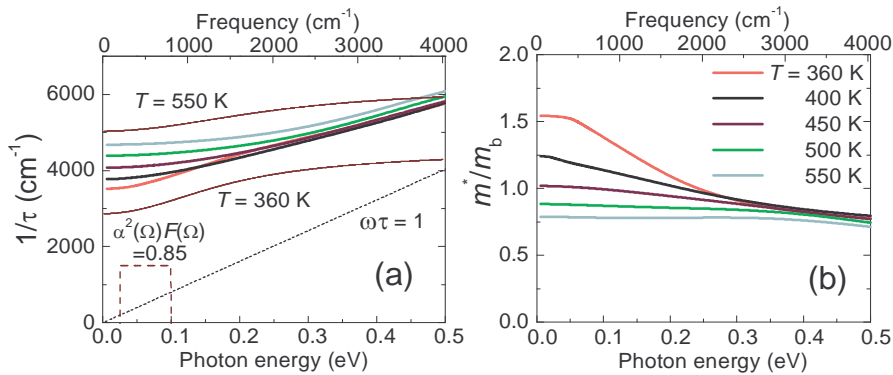,width=120mm,height=50mm}
\caption{(color online) (a)
The scattering rate $1/\tau$ and (b) the mass enhancement factor
$m^*/m_b$ in the metallic state of VO$_2$ are plotted as a
function of frequency for different temperatures (labelled in
panel (b)). Extended Drude analysis was performed after phonons
were removed from the conductivity in the frequency range 0.04 -
0.08 eV. In panel (a), the dotted line satisfies the equation
$\omega\tau$=1, the dashed line is a model phonon spectral density
$\alpha^2(\Omega)F(\Omega)$ (see text for details), and thin solid
lines are the electron-phonon scattering rates at $T$ = 360 K and
$T$ = 550 K calculated from $\alpha^2(\Omega)F(\Omega)$.}
\label{xtendedDrude}
\end{figure*}

Next, we perform an extended Drude analysis to obtain the
frequency and temperature dependence of the effective scattering
rate (1/$\tau$) and mass enhancement factor ($m^*/m_b$) of the
charge carriers.\cite{basovreview,footnote2} These are plotted in
Fig.~\ref{xtendedDrude}a,b.

\begin{align}
\frac{1}{\tau(\omega)}=\frac{\omega_p^2}{4\pi}\frac{\sigma_1(\omega)}{\sigma_1^2(\omega)+\sigma_2^2(\omega)}\\
\frac{m^*(\omega)}{m_b}=\frac{\omega_p^2}{4\pi\omega}\frac{\sigma_2(\omega)}{\sigma_1^2(\omega)+\sigma_2^2(\omega)}
\label{xtendDrude}
\end{align}

In the above equations, $\sigma_2(\omega)$ is the imaginary part
of the conductivity and $m_b$ is the band mass of the charge
carriers. We find that the scattering rate (in the zero frequency
limit) increases linearly with temperature and thereby leads to
the corresponding increase of resistivity (see Fig.~\ref{sigmaT}b
and inset). The absolute value of the scattering rate at all
measured temperatures and infrared frequencies is higher than the
excitation energy ($\hbar\omega$). The canonical criterion for
well-defined quasi-particles is that $1/\tau$ $\ll$
$\omega$.\cite{allencomments,basovreview} The scattering rate in
VO$_2$ certainly violates this criterion in common with the high
$T_c$ cuprates.\cite{basovreview} P. B. Allen has defined the
criterion for the validity of the Boltzmann equation and Fermi
liquid theory as $\hbar/\tau \ll E_F$ (or $N(0)\hbar/\tau \ll 1$)
where $E_F$ is the Fermi energy and $N(0)$ is the density of
states at the Fermi level.\cite{allencomments} In
Fig.~\ref{xtendedDrude}a, we see that $\hbar/\tau \approx$ 0.5 eV
which is comparable to $E_F$ $\approx$ 0.5 eV.\cite{georges}
Alternatively, taking $N(0)$= 4.16/eV molecule,\cite{allenprl} we
find that $N(0)\hbar/\tau$ is of order unity. Hence, charge
carriers in metallic VO$_2$ are not well-defined quasiparticles.

In Fig.~\ref{xtendedDrude}b we see that the mass enhancement at
$T$=360 K is a strong function of frequency. The frequency
dependence of $m^*/m_b$ becomes weaker at higher temperatures.
Also, $m^*/m_b$ in the low frequency limit increases significantly
with decreasing temperature i.e. as one approaches the insulating
state. Mass enhancement has been predicted in a strongly
correlated electron gas in the vicinity of a transition to the
Mott insulator.\cite{brinkman} In VO$_2$, the absolute value of
$m^*/m_b$ is modest at $T$=360 K consistent with recent
photoemission results.\cite{okazaki} The increase in $m^*/m_b$
with temperature on approaching the insulating state is
nevertheless significant and provides evidence for correlations in
VO$_2$. However, the mass enhancement is not as spectacular as
seen with doping in Sr$_{1-x}$La$_x$TiO$_{3}$, a classic system
that exhibits a mass-diverging-type Mott transition.\cite{tokura}
Several correlated systems are known to avoid mass-divergence near
the MIT.\cite{imada-rmp,padilla}

We now investigate the metallic behavior of VO$_2$ in light of the
IRM criterion. At high temperatures, the resistivity of typical
metals and alloys saturates at values which give mean free paths
($l$) of the quasi-particles that are comparable to the lattice
spacing ($d$).\cite{allencomments} A complementary viewpoint is
that resistivity saturates when $l$ becomes comparable to the de
Broglie wavelength $\lambda_F$ ($\lambda_F=2\pi/k_F$, where $k_F$
is the Fermi momentum).\cite{emery} The real part of the optical
conductivity in the dc limit $\omega \rightarrow 0$
(Fig.~\ref{sigma1}) can be used to estimate the mean free path of
the electrons in the metallic phase of VO$_2$ from the equation
$\rho=\frac{3\pi^2\hbar}{e^2k_F^2l}$.\cite{gunnarsson} Here,
$\rho$ is the $dc$ resistivity ($\rho = 1/\sigma_1(\omega
\rightarrow 0$)). Since a single $d$-electron per vanadium atom is
delocalized,\cite{allenprl,georges,softxray} one obtains $k_F
\approx$ 10$^8$ cm$^{-1}$ assuming a spherical Fermi surface for
simplicity. Thus, the mean free path is $\approx$ 2.8 \AA~at $T$ =
360 K i.e. $l$ is nearly equal to the lattice spacing (2.85 \AA)
and less than $\lambda_F$ ($\approx$ 6 \AA).\cite{footnote3}
Therefore, the Bloch-Boltzmann transport model no longer applies
and the quasiparticle concept breaks down. It is observed that
departure from the Boltzmann transport theory typically occurs for
$l < 10$ \AA~in metals.\cite{allenprb} If VO$_2$ were to avoid
breakdown of Boltzmann transport through resistivity saturation
similar to common metals, then saturation is expected to occur
already at $T$ = 360 K (Fig.~\ref{sigmaT}b). Instead, resistivity
increases continuously beyond the IRM limit.

It has been argued that a quantum theory of electrons which also
includes correlation effects is required to give a correct account
of resistivity behavior at short values of the electronic mean
free path.\cite{gunnarsson} Within this theoretical framework, the
optical conductivity of a correlated metal provides important
insights into saturation phenomena.  The low energy response of
correlated metals at reduced temperatures often reveals a
Drude-like mode at far-IR frequencies attributable to well-defined
charge carriers followed by a broad, incoherent part at higher
energies. With increasing $T$, the coherent Drude weight is
transferred to the incoherent part.\cite{imada-rmp} Resistivity
saturation is expected to occur when there is no further coherent
spectral weight to be shed and only the incoherent part remains.
In this model, the limiting $dc$ conductivity $\sigma_s$(0) of the
incoherent part (corresponding to resistivity saturation) can be
estimated from the $f$-sum rule:

\begin{equation}
\frac{\sigma_s(0)W}{\gamma}=\hbar\int_{0}^{\omega_c}\sigma_1(\omega)d\omega
\label{incoherent}
\end{equation}

Here $W$ is the one-particle bandwidth and $\gamma$ is a weighting
factor ($\approx$ 2) depending upon the shape of the incoherent
part of the conductivity.\cite{gunnarsson} We integrate the
optical conductivity in the metallic phase of VO$_2$ up to the
frequency of the minimum in the conductivity, beyond which
inter-band transitions are clearly present. We take $W =$ 2.59 eV
from recent LDA calculations.\cite{georges} Then using
eq.~\ref{incoherent}, we obtain $\sigma_s(0)$ $\approx$ 1200
$\Omega^{-1}$cm$^{-1}$. We will refer to this lower limit of the
conductivity (upper limit of resistivity) as the ``$f$-sum rule
limit." Note that the $f$-sum rule resistivity is higher than the
IRM resistivity (see Fig.\ref{sigmaT}a,b). The conductivity
decreases with increasing temperature but no saturation is
observed because $\sigma_1(\omega \rightarrow 0)$ at $T$ = 550 K
is still above the $f$-sum rule limit. The coherent part of the
conductivity (the Drude peak at $\omega$ = 0) persists beyond the
IRM limit in a regime where the quasiparticle picture is no longer
valid. This is remarkable because some other bad metals with
exceptionally short mean free paths usually exhibit a peak at
$finite$ frequency in $\sigma_1(\omega)$.\cite{hussey,takenaka}

At this stage, we briefly discuss the effect of phonons on the
electronic properties of metallic VO$_2$. We calculate 1/$\tau$
for electron-phonon scattering using a square shaped Eliashberg
function $\alpha^2(\Omega)F(\Omega)$ that mimics the phonon
spectral density of VO$_2$.\cite{basovreview}

\begin{align}
\frac{1}{\tau}(\omega,T)=
\frac{\pi}{\omega}\int_0^{\infty}d\Omega\alpha^2(\Omega)F(\Omega)
\Big[2\omega\coth\Big(\frac{\Omega}{2T}\Big)
\nonumber\\-(\omega+\Omega)\coth\Big(\frac{\omega+\Omega}{2T}\Big)+(\omega-\Omega)\coth\Big(\frac{\omega-\Omega}{2T}\Big)\Big]
\label{elph}
\end{align}

We note that the precise form of $\alpha^2(\Omega)F(\Omega)$ is
not known for VO$_2$ but the shape of the calculated 1/$\tau$
plots is not very sensitive to it. The upper cutoff for
$\alpha^2(\Omega)F(\Omega)$ is taken as 0.1 eV which is the
maximum phonon energy of VO$_2$.\cite{barker} As shown in
Fig.~\ref{xtendedDrude}a, electron-phonon coupling theory predicts
1/$\tau$ to be independent of frequency and strongly temperature
dependent for frequencies greater than the maximum phonon
energy.\cite{basovreview} However in VO$_2$, 1/$\tau$ is frequency
dependent and nearly temperature independent at frequencies well
above the maximum phonon energy (Fig.~\ref{xtendedDrude}a).
Spectroscopic evidence and the lack of resistivity saturation
indicate that electron-phonon scattering plays a less significant
role in charge dynamics in the metallic phase of VO$_2$ compared
to electronic correlations. Therefore, the linear increase of
resistivity and 1/$\tau$ ($\omega \rightarrow$ 0) with temperature
cannot be due to electron-phonon scattering.

We note that there is broad consensus on the breakdown of the
conventional quasiparticle picture for $l \lesssim d$ (or $l
\lesssim \lambda_F$).\cite{allencomments,emery,gunnarsson} Despite
attempts to understand transport in bad metals, including
proposals beyond conventional Fermi liquid theory, there is no
convergence of
views.\cite{emery,gunnarsson,merino,oganesyan,anderson} On the
other hand, one cannot rule out the possibility of electronic
phase segregation on a length scale smaller than infrared
wavelengths. Also, silver films grown under certain conditions are
known to exhibit bad metal behavior due to their peculiar
microstructure and not due to any novel physics.\cite{hebard}
Nevertheless, empirical evidence indicates that a common feature
of bad metals is the predominance of electronic correlations in
their transport properties due to proximity to a Mott insulating
phase.\cite{hussey} We find that metallic VO$_2$ fits into this
picture and suggest that the on-site electronic Coulomb repulsion
is crucial to bad metal behavior and the MIT even though the
insulating state may not be a conventional Mott
insulator.\cite{georges,softxray}

\section{Conclusion}

We have determined the optical constants of VO$_2$ films in the
insulating and metallic states over a wide temperature and
frequency range. We find that VO$_2$ is a correlated metal above
$T_c$ and the charge dynamics are dominated by electronic
correlations rather than electron-phonon scattering. In the
metallic state, the length and energy scales associated with the
scattering processes make it problematic to describe the charge
carriers within Fermi liquid theory. However, in contrast to some
other bad metals,\cite{hussey,takenaka} the Drude peak persists in
VO$_2$ for $l \lesssim d$. We establish that in metallic VO$_2$,
the increase in resistivity with temperature beyond the IRM limit
is due to the increase of scattering rate of charge carriers
rather than a decrease of carrier density. The data is consistent
with the model based on the fundamental $f$-sum rule in optics in
which resistivity saturation is not expected at the IRM
limit.\cite{gunnarsson}

\section{Acknowledgments}

We thank G. Kotliar, S. A. Kivelson, O. Gunnarsson, T. Tiwald, C.
Marianetti, V. Oganesyan, and Zhiqiang Li for discussions. This
work was supported by Department of Energy Grant No.
DE-FG03-00ER45799 and by ETRI.


\begin{references}
\bibitem{footnote1} Strong sapphire substrate phonon features prevent a
precise determination of lineshapes of VO$_2$ phonons. This does
not affect the results contained in this paper.
\bibitem{footnote2} The experimentally determined $\omega_p$ = 2.75 eV was used for the extended
Drude analysis. The conclusions of this paper will not be affected
if a higher value of $\omega_p$ from band theory is used instead.
\bibitem{footnote3} For a Fermi surface with $M$ sheets, the mean
free path is reduced by a factor $M^{1/3}$
(Ref.\cite{gunnarsson}).
\end{references}
\end{document}